\documentclass{elsart}
\usepackage{lineno,hyperref}
\modulolinenumbers[5]

\journal{Journal of \LaTeX\ Templates}

\usepackage{color} 
\usepackage{ulem} 
\usepackage{placeins}
\usepackage{ulem} 
\usepackage{booktabs}
\usepackage{blindtext}
\makeatletter
\newcommand{\colorcaption}[2][]{%
  \begingroup%
  \renewcommand{\@caption@fignum@sep}{ (Color online). }%
  \caption[#1]{#2}%
  \endgroup%
}
\makeatother
\usepackage{amssymb}
\usepackage{graphicx}
\usepackage{rotating}
\usepackage{tikz}
\begin{document}



\begin{frontmatter}
\title{Shell-model study for GT-strengths corresponding to $\beta$ decay of $^{60}$Ge and $^{62}$Ge}
\author{Vikas Kumar$^{1}$, Anil Kumar$^{2}$ and Praveen C. Srivastava$^{2}$}
\address{$^{1}$Department of Physics, Institute of Science, Banaras Hindu University, Varanasi 221005, India}
\address{$^{2}$Department of Physics, Indian Institute of Technology, Roorkee
247 667, India}


\date{\today}

\begin{abstract}  
In the present work, we have reported a comprehensive shell-model study of GT-strengths for recently available experimental data for $^{60}$Ga and $^{62}$Ga from RIKEN Nishina Center [Phys. Rev. C 103, 014324 (2021)] populated by $\beta$ decay of the $^{60}$Ge and $^{62}$Ge, respectively. We have performed shell-model calculations in two different model spaces, the first set of calculations in the $fp$ model space using KB3G and GXPF1A interactions, while the second set  in  $f_{5/2}pg_{9/2}$ model space using JUN45 and jj44b effective interactions. Our shell-model results in $fp$ model space are in a reasonable agreement with the available experimental data.  

\end{abstract}


\end{frontmatter}

\section{Introduction}
The study of nuclei away from the stability line is one of the major
goals in nuclear physics \cite{stroberg,nunes,otsuka}. It is possible
to extract several useful information about nuclear structure using the
study of Gamow-Teller (GT) transition strengths. The GT-transition is
the weak interaction process of spin-isospin ($\sigma \tau $) type in
nuclei \cite{fujita}. The study of GT-strengths is also important for
the astrophysics process, and it is responsible for electron capture
during the core collapse of supernovae. The experimental B(GT)
strengths can be obtained using $\beta $-decay and charge-exchange (CE)
reactions. One can only access the states lower than the decay
$Q$-value in the $\beta $-decay, while it is possible to measure
GT-transitions at higher energies using CE reactions such as $(p,n)$,
$(n,p)$, ($d$,\,$^{2}$He), or ($^{3}$He,\,$t$). A recent
review of the experimental and theoretical status on the single and
double beta decays is available in \cite{jouni_review}.

There are several recent experimental data of GT-strengths available for
the $fp$ shell nuclei. The observation of the $\beta $-delayed
$\gamma $-proton decay of $^{56}$Zn and its impact on the Gamow-Teller
strength evaluation is reported in Ref.~\cite{Orrigo1}. Also, they observed
evidence for fragmentation of the isobaric analogue state (IAS) in
$^{56}$Cu due to isospin mixing. The results of the $\beta $ decay of three
proton-rich nuclei with $T_{z}=-2$, namely $^{48}$Fe, $^{52}$Ni, and
$^{56}$Zn were reported in \cite{Orrigo2}. The beta decay results of
$f_{7/2}$ nuclei $^{54}$Ni, $^{50}$Fe, $^{46}$Cr, and $^{42}$Ti produced
in fragmentation reactions at GSI are reported in Ref.~\cite{Molina}, further,
these results are compared with the charge exchange reaction results corresponding
to $T z=+1$ to $T z=0$ performed at RCNP-Osaka. Adachi et al. were
performed a high-energy-resolution ($^{3}$He, t) CE reaction experiment
on $T_{z}=1$ nuclei $^{46}$Ti and $^{54}$Co for the study of GT transition
strengths to the daughter $T_{z}=0$ nuclei $^{46}$V and $^{54}$Co, respectively
\cite{Adachi1,Adachi2}.

Apart from the above, several other experiments have been carried out
to investigate the GT strength for the $\beta ^{+}$ decay of
$^{60,62}$Ge, some are direct $\beta $ decay measurements while others
are nuclear reactions based. The $\beta $ decay of $^{60}$Ge was
measured at NSCL, MSU, and found with a $100\%$ branching fraction and
half-life $T_{1/2}=20^{+7}_{-5}$ ms \cite{Ciemny2016}. Another
experiment was performed to study the $\beta $ decay of $^{62}$Ge at
GANIL and determined the half-life (129 $\pm $ 35) ms \cite{Lopez2001}.
The $\beta $ decay properties of the nucleus $^{60}$Ga were measurement
for the first time in \cite{Mazzocchi2001} by using the
fusion-evaporation reaction $^{28}$Si($^{36}$Ar, p3n). The study of the
GT strength for the $\beta $ decay of the $T=1$, $J^{\pi }=0$ ground
state of $^{62}$Ge into the different excited states of the odd-odd
$N=Z$ nucleus $^{62}$Ga is given in \cite{Grodner}. In this work, they
have measured total six excited states of $^{62}$Ga, below 2.5 MeV
through $\beta $ decay. In Ref.~\cite{Kucuk2017}, they have measured
total 16 $\beta $ decay transitions of neutron-deficient nuclei ranging
from chromium to germanium with $T_{z}=-1/2$ and $-1$. The rotational
structure of $T=0$ and $T=1$ bands in the $N=Z$ nucleus $^{62}$Ga using
shell model and cranked Nilsson-Strutinsky model is studied in
\cite{Aberg}. The spectroscopy of the $^{62}$Ga nucleus is studied for
the high-spin states using the shell-model and deformed shell-model
given in Ref.~\cite{PCS}. The role of neutron-proton pairing and its
impact on GT-transitions for $A=42-48$ nuclei is reported by Pittel et
al. in Ref.~\cite{Pittel}. In Ref.~\cite{Pittel}, it was observed that
the varying strength parameters for the isoscalar and isovector showed
different but systematic effects on GT-transition properties and also
on the corresponding energy spectra.

Recently, an experiment has been performed at RIKEN in which they have
populated several new excited states and $\beta $-feeding branching fractions
corresponding to these states of $^{60,62}$Ga \cite{Orrigo}. In this experiment
decay schemes, absolute Gamow-Teller and Fermi transition strengths have
been determined for the $\beta ^{+}$ decay of $^{60,62}$Ge. Also, they
have improved the precision in the half-lives of $^{62}$Ge [73.5(1) ms],
$^{60}$Ge [25.0(3) ms] and $^{60}$Ga [69.4(2) ms] from the literature
\cite{Ciemny2016,Kucuk2017,Mazzocchi2001,Grodner}. New information has
been reported for the energy levels of $^{60}$Ga and on the
$1/2^{-}$ a first excited state in $^{59}$Zn. Also, evidences of populations
of levels in four nuclei $^{60}$Ga, $^{60}$Zn, $^{59}$Zn, and
$^{59}$Cu in the decay chain of $^{60}$Ge are reported. These new experimental
data for the $\beta ^{+}$ decay of $^{60,62}$Ge motivated us to perform
the shell-model calculations for the GT-strengths and analyze the nuclear
structure.

In the present work, our aim is to explain the recently available
experimental data \cite{Orrigo} for the Gamow-Teller transition
strengths corresponding to $^{60}$Ge($0^{+}$) $\rightarrow $
$^{60}$Ga($1^{+}$) and $^{62}$Ge($0^{+}$) $\rightarrow $
$^{62}$Ga($1^{+}$) transitions in the framework of the nuclear
shell-model. We have used two different model spaces such as $fp$ and
${f_{5/2}pg_{9/2}}$, using GXPF1A and KB3G interactions for $fp$ shell
and JUN45 and jj44b for $f_{5/2}pg_{9/2}$ shell, respectively. First,
we have calculated the energy spectra for low-lying states in order to
test the predictive power of our computed wave functions and found the
energy spectra are in good agreement with the available experimental
data. After that, we have used those wavefunctions to compute
GT-strengths.

This paper is organized as follows. In Sec.~\ref{gt}, we give a short
overview of GT-formalism. Results and discussions about energy spectra
and the GT-strengths are presented in Sec.~\ref{results}. Finally, in
Sec.~\ref{conclusions} the summary and conclusions are discussed.


\section{Theoretical Formalism}\label{gt}

In the present work, we have performed theoretical calculations using the nuclear shell-model. The Hamiltonian contains single-particle energy  and a  two-body matrix elements (TBMEs). The shell-model Hamiltonian can be written as 

\begin{equation}
~~~~~~~~~~~~	H = T + V = \sum_{\alpha}{\epsilon}_{\alpha} c^{\dagger}_{\alpha} c_{\alpha} + \frac{1}{4} \sum_{\alpha\beta \gamma \delta}v_{\alpha \beta \gamma \delta} c^{\dagger}_{\alpha} c^{\dagger}_{\beta} c_{\delta} c_{\gamma},
\end{equation}
where $\alpha = \{n,l,j,t\}$ is a single-particle state and the corresponding single-particle energy is $\epsilon_{\alpha}$. $c^{\dagger}_{\alpha}$ and $c_{\alpha}$ are the creation and annihilation operators, respectively. 
$v_{\alpha \beta \gamma \delta} = \langle\alpha \beta | V | \gamma \delta\rangle $ are the antisymmetrized TBMEs.

The Gamow-Teller transition strength B(GT) is calculated using  the following
expression,
\begin{equation}
~~~~~~~~~~~~ B(GT ; i \rightarrow f ) = q^2\frac{1}{2J_i + 1}  \, |{\langle {f}|| \sum_{k}{\sigma^k\tau_{\pm}^k} ||i \rangle}|^2,
\end{equation}
where  $\tau_+|p\rangle = |n\rangle$ , $\tau_-|n\rangle  = |p\rangle$, the index $k$ runs over the single-particle orbitals, 
$|i \rangle$ and $|f \rangle$ describe the states of the parent and daughter nuclei, respectively.
 The $q$ is the quenching factor.

In the beta decay, the $ft$ value corresponding to $GT$ transition from the initial state $i$ of the parent nucleus to the final state $f$
in the daughter nucleus is expressed as 

\begin{equation}
~~~~~~~~~~~~~~~~~~~~  f_{\rm A}t_{i\rightarrow{f}} = \frac{6177}{(g_{\rm A}^{\rm free})^2[ B(GT;i\rightarrow{f})]},
\end{equation}
where B(GT) is the Gamow-Teller  transition strength,  $g^{\rm free}_{\rm A}$  is the free-nucleon value of the axial-vector coupling constants and  $f_{\rm A}$ is  the axial-vector phase space
factor that  contains the lepton kinematics.  
We have calculated the phase space factor $f_{\rm A}$ using the  parameters given by
Wilkinson and Macefield \cite{wilkinson} together with the correction factors given in Refs. \cite{Sirlin,wga}. 

The total half-life $T_{1/2}$  is related to  the partial half-life as 
\begin{equation}
~~~~~~~~~~~~~~~~~~~~~~~~~ \frac{1}{T_{1/2}}= {\sum_f {\frac{1}{t_{i\rightarrow{f}}}}},
\end{equation}
where $f$ runs over all the possible daughter states that are populated through GT transitions.

 We have also computed the $Q$ values in this work, which can be defined as
\begin{equation}
~~~~~~~~~~~Q = [(E(SM)_i + E(C)_i) - (E(SM)_f + E(C)_f)],
\end{equation}
where  $E(SM)$ is the nuclear binding energy of the interaction of the valence particles
among themselves calculated using shell-model, the valence space Coulomb energy is given by $E(C)$ \cite{Caurier59}, 
 and subscripts $i$ and $f$ denote the parent and daughter nuclei, respectively.

A comprehensive study of GT-strengths for $sd$ and $fp$ shell nuclei is reported in Refs. \cite{brownwild,Mart}. In our previous work, we have reported GT-strengths for $sd$ and $fp$ shell nuclei in
Refs. \cite{Archana2018,Anil2020,Vikas2016,Vikas22016,Anil2020PRC,Anil2021EPJA,PriyankaPRC}.
The shell-model results for high-spin states and band terminations in $^{67}$As 
we have recently reported  in Ref. \cite{vikasnpa} using JUN45 and jj44b effective interactions.


\begin{figure}[h]
\begin{center}
	\includegraphics[width=13.5cm,height=10cm]{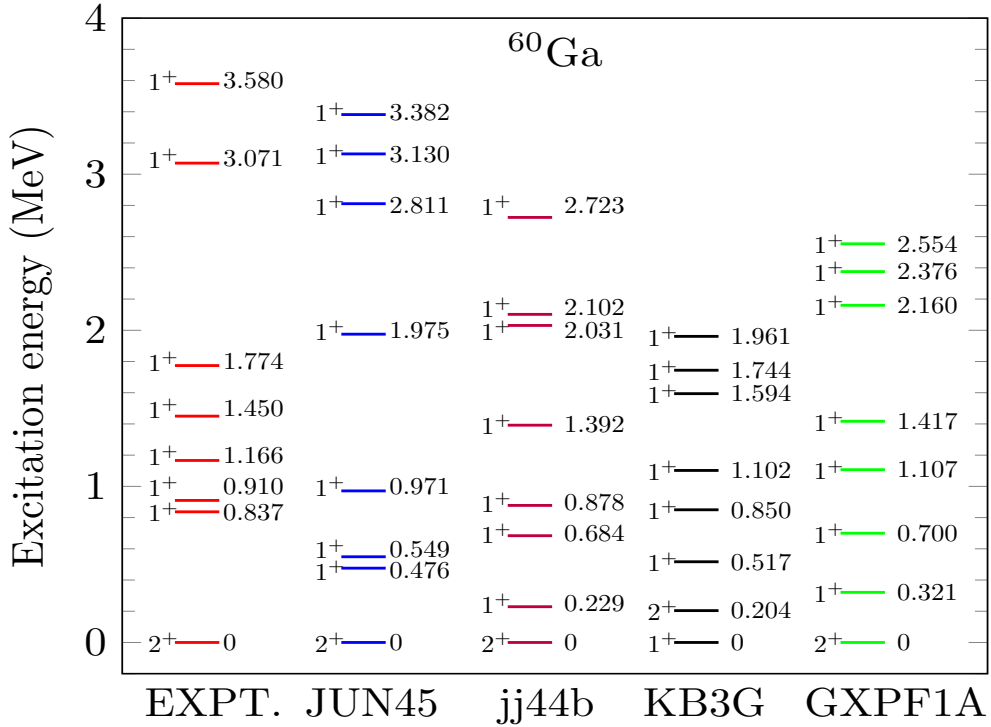}
	\caption{Comparison of energy levels between calculated and experimental data for $^{60}$Ga.}
	\label{60ga}
	\end{center}
\end{figure}

\begin{figure}[h]
\begin{center}
	\includegraphics[width=13.5cm,height=10cm]{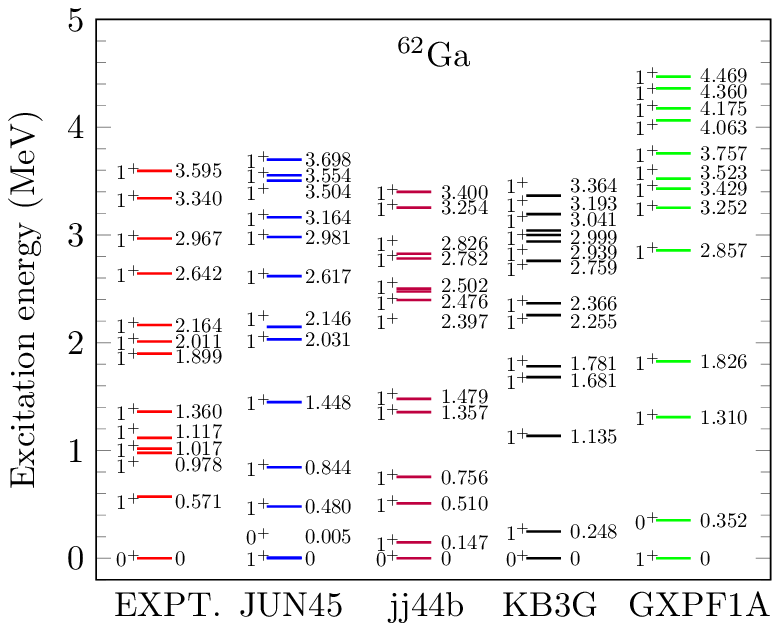}
	\caption{Comparison of energy levels between calculated and experimental data for $^{62}$Ga.}
	\label{62ga}
	\end{center}
\end{figure}


\section{Results and Discussions}\label{results}

\subsection{ {\bf  Adopted model space and Hamiltonian}}
    In this work, we have performed shell-model calculations for two different model spaces i.e. in $f_{5/2}pg_{9/2}$ and $fp$ spaces. In $f_{5/2}pg_{9/2}$ model space, we  have used two different effective interactions JUN45 \cite{jun45} and jj44b \cite{jj44b}, while in $fp$  model space  used KB3G \cite{KB3G} and GXPF1A \cite{GXPF1A} effective interactions. The $^{56}$Ni is taken as the inert core with the spherical orbits $1p_{3/2}$, $0f_{5/2}$, $1p_{1/2}$, and $0g_{9/2}$ in $f_{5/2}pg_{9/2}$ model space, while in $fp$ space, the $^{40}$Ca is taken as inert core with the spherical orbits $0f_{7/2}$, $1p_{3/2}$, $0f_{5/2}$,  and $1p_{1/2}$. 
    
 The JUN45 interaction is established from Bonn-C potential, further, the  single-particle energies and two-body matrix elements of JUN45 interaction were modified empirically in the mass region A = 63$\sim$69. 
 The jj44b interaction was fitted with 600 experimental binding energies and excitation energies from nuclei with $Z = 28-30$ and $N = 48-50$. The 30 linear combinations of $JT$ coupled two-body matrix elements (TBME) are varied and giving the rms deviation of about 250 keV from  the experiment. 
 
        
        The KB3G interaction is based on KB3 \cite{KB3}, the mass dependence and original monopole changes employed into KB3 to get a new version KB3G. The idea behind this was to  treat properly the $N = Z = 28$ shell closure and its surroundings. 
        
        
 
        The modified GXPF1 \cite{GXPF1} interaction is referred to as GXPF1A interaction, GXPF1 is derived from the Bonn-C potential. Further, the 70 well-determined linear combinations of 4 single-particle energies and 195 two-body matrix elements are modified by iterative fitting calculations to about 700 experimental energy data out of 87 nuclei. 
        The shell-model interactions which we have taken in the present study are  isospin symmetric, thus they will give the same results for mirror nuclei.

     Further, to see the importance of lower  $f_{7/2}$ orbit, we have also reported the shell-model results in $fp$ valence space  corresponding two different effective interactions KB3G and GXPF1A. Due to the huge dimensions of the energy matrix, we have imposed truncation to protons/neutrons in $fp$ space. For the case of $\beta^+$ decay of $^{60}$Ge, we have fixed the minimum six protons in the $f_{7/2}$ orbital while in the case of $^{62}$Ge we have fixed minimum six protons/neutrons and two in $f_{7/2}$  and  $p_{3/2}$ orbitals, respectively.
       We did not impose any truncation on nucleons for shell-model calculations in $f_{5/2}pg_{9/2}$ model space.   The shell-model calculations are performed using the code NuShellX@MSU \cite{NuShellX@MSU} and KShell \cite{kshell}.  The energy spectra corresponding to $^{60}$Ga and  $^{62}$Ga  are shown in Figs. \ref{60ga} and  \ref{62ga}, respectively.

\begin{sidewaystable}
	\begin{center}
		\caption{\label{tab:summary1} Present table shows initial and final nuclei, 
			the number of GT transitions, transitions up to the excitation energy in MeV and the references are 
			given in the last column for comparison with the theoretical results.}
		\begin{tabular}{| p {1.5 cm} | p {1.5 cm} | p {2 cm} | p {1.5 cm} | p {1.5 cm} | p {1.5 cm} | p {1.5 cm} | p {1.5 cm} | p {0.5 cm} |}
			\hline
			\hline
			Initial   & Final & Transitions (No.) &    EXPT.    & JUN45  & jj44b  &  KB3G  &    GXPF1A &   Ref.  \\
			\hline    
			$^{60}$Ge($0^{+}$)  & $^{60}$Ga($1^{+}$)   &   7 & 3.580 & 3.382 &  2.723 &  1.961 &2.554 & \cite{Orrigo}\\
			$^{62}$Ge($0^{+}$)  & $^{62}$Ga($1^{+}$)   &  12 & 3.595 & 3.698 &  3.400  & 3.364 &4.469 & \cite{Orrigo}\\
			\hline
		\end{tabular}
	\end{center}
\vspace{0.5cm}
	\begin{center}
		\caption{\label{tab:halflife} Comparison of the theoretical Q values and $\beta$-decay half-lives with the experimental data for the concerned transitions.}
		\begin{tabular}{| p {3.5 cm} | p {2.1 cm} | p {1.2 cm} | p {1 cm} | p {1 cm} | p {1.5 cm} | p {1.2 cm} | p {1.2 cm} | p {1 cm} | p {1 cm} | p {1.5 cm} |}
			\hline
			\hline
			Process  &      \multicolumn{5} {c|} {Q (MeV)}  & \multicolumn{5} {c|} {Half-life (ms)}\\
			\cline{2-6}
			\cline{6-11}
			& Expt.   & JUN45 &  jj44b  & KB3G & GXPF1A &EXPT. & JUN45 & jj44b&  KB3G  &  GXPF1A  \\
			\hline  
			$^{60}$Ge$(0^+)$$\rightarrow$$^{60}$Ga$(1^+)$  & 12.338(27)   & 14.904 &  14.396 &  12.907 & 14.220  & 25.0(3)   & 28.2 & 23.5 & 42.7 & 87.9 \\ 
			$^{62}$Ge$(0^+)$$\rightarrow$$^{62}$Ga$(1^+)$  & 9.730(140$^\#$)  & 10.839 &  10.839  & 11.072 & 11.074 & 73.5(1) & 117.8 & 112.1 & 417.9 & 576.0 \\ 
			\hline
		\end{tabular}
	\end{center}

\end{sidewaystable}

\begin{sidewaystable}
	\begin{center}
		\caption{\label{tab:GT} Comparison of the theoretical excitation energies with the experimental data for the concerned transitions together with the quenched GT strengths.}
		\begin{tabular}{| p {1.5 cm} | p {1.6 cm} | p {1.2 cm} | p {1.2 cm} | p {1.2 cm} | p {1.2 cm} | p {1.5 cm} | p {1.3 cm} | p {1.2 cm} | p {1.2 cm} | p {1.2 cm} | p {1.5 cm} | p {0.5 cm} |}
			\hline
			\hline
			$^{A}Z_{i}(J^{\pi})$ & $^{A}Z_{f}(J^{\pi})$  &   \multicolumn{5} {c|} {$E_{x}$ Energy (MeV)}  & \multicolumn{5} {c|} {$B(GT)$}& Ref.\\
			\cline{3-7}
			\cline{8-12}
			&                     &  EXPT.& JUN45 &  jj44b  & KB3G & GXPF1A &EXPT. & JUN45 & jj44b&  KB3G  &  GXPF1A &  \\
			\hline  
			$^{60}$Ge$(0^+)$  & $^{60}$Ga$(1_1^+)$  &  0.837 & 0.476 &   0.229   &0.000    &0.321   & 0.11(3)   &0.0053 & 0.0087 &0.0265  & 0.0636 &\cite{Orrigo}\\
			& $^{60}$Ga$(1_2^+)$  &  0.910 & 0.549 &   0.684   & 0.517   &0.700   & 0.044(6)                     &0.2830 & 0.0836 &0.5423  & 0.0010 &             \\
			& $^{60}$Ga$(1_3^+)$  & 1.166 & 0.971 &   0.878   & 0.850   &1.107  & 0.074(9)                     &0.3512 & 1.1962 &0.0066  & 0.2339 &             \\
			& $^{60}$Ga$(1_4^+)$  & 1.450 & 1.975&  1.392   & 1.102  &1.417  & 0.11(1)                      &1.3280 & 0.2629 &0.1153  & 0.1413 &             \\
			& $^{60}$Ga$(1_5^+)$  & 1.774 & 2.811&  2.031   & 1.594  &2.160  & 0.11(1)                      &0.0029 & 0.3826 &0.2024  & 0.0013&             \\
			& $^{60}$Ga$(1_6^+)$  & 3.071 & 3.130&  2.102   & 1.744  &2.376  & 0.18(2)                      &0.0026 & 0.0024 &0.1894  & 0.1767 &             \\
			& $^{60}$Ga$(1_7^+)$  & 3.580 & 3.382&  2.723   & 1.961  &2.554  & 0.14(1)                      &0.5885 & 0.5417 &0.0048  & 0.0685 &             \\
\hline			
			$^{62}$Ge$(0^+)$ & $^{62}$Ga$(1_1^+)$   &0.571   &0.000  &0.147      & 0.248   & 0.000  &  0.068(6)  &0.0290 &0.1904  & 0.0000 & 0.0000 &\cite{Orrigo}\\
			& $^{62}$Ga$(1_2^+)$   &0.978   &0.480  &0.510      & 1.135  & 1.310 &  0.047(4)                   &1.0806 &0.3730  & 0.3777 & 0.3631 &             \\
			& $^{62}$Ga$(1_3^+)$   &1.017  &0.844  &0.756      & 1.681  & 1.826 &  0.067(6)                   &0.0429 &0.6114  & 0.1174 & 0.0000 &             \\
			& $^{62}$Ga$(1_4^+)$   &1.117  &1.448 &1.357     & 1.781  & 2.857 &  0.011(2)                   &0.0466 &0.2142  & 0.0017 & 0.0541 &             \\
			& $^{62}$Ga$(1_5^+)$   &1.360  &2.031 &1.479     & 2.255  & 3.252 &  0.022(2)                   &0.0053 &0.0086  & 0.0024 & 0.0060 &             \\
			& $^{62}$Ga$(1_6^+)$   &1.899  &2.146 &2.397     & 2.366  & 3.429 &  0.025(3)                   &0.0183 &0.0098  & 0.0043 & 0.0809 &             \\
			& $^{62}$Ga$(1_7^+)$   &2.011  &2.617 &2.476     & 2.759  & 3.523 &  0.045(5)                   &0.0032 &0.0168  & 0.0859 & 0.0213 &             \\
			& $^{62}$Ga$(1_8^+)$   &2.164  &2.981 &2.502     & 2.939  & 3.757 &  0.13(1)                    &0.0027 &0.0036  & 0.0054 & 0.0067 &             \\
			& $^{62}$Ga$(1_9^+)$   &2.642  &3.164 &2.782     & 2.999  & 4.063 &  0.029(7)                   &0.0037 &0.0058  & 0.0080 & 0.0027 &             \\
			& $^{62}$Ga$(1_{10}^+)$&2.967  &3.504 &2.826     & 3.041  & 4.174 &  0.028(5)                   &0.0133 &0.0070  & 0.0009 & 0.0017 &             \\
			& $^{62}$Ga$(1_{11}^+)$&3.340  &3.554 &3.254     & 3.193  & 4.360 &  0.030(7)                   &0.0001 &0.0051  & 0.0166 & 0.0000 &             \\
			& $^{62}$Ga$(1_{12}^+)$&3.595  &3.698 &34.00     & 3.364  & 4.469 &  0.070(1)                   &0.0732 &0.0012  & 0.0004 & 0.0000 &             \\
			\hline
		\end{tabular}
	\end{center}
\end{sidewaystable}


For the further calculations of the $B(GT)$ and summed $B(GT)$ values, we have  used the quenching factors in axial-vector coupling constants from our previous work \cite{Vikas2016}, these values are q=0.684 $\pm$ 0.015 for JUN45 and jj44b, while q = 0.660 $\pm$ 0.016, for KB3G and GXPF1A interactions.

In Table \ref{tab:summary1}, we have reported the number of GT transitions  up to the excitation energy in MeV
 corresponding to experimental data and theoretical results. In  Table \ref{tab:halflife}, we have reported a comparison of the theoretical $\beta$-decay half-lives 
 with the experimental data for the concerned transitions.  For the calculations of half-lives, we have  used the experimental $Q$-values \cite{Orrigo}. 
 We have also calculated $Q$ values from the shell-model, and corresponding results are presented in this Table. 
 The comparison of the theoretical excitation energies with the experimental data for the concerned transitions together
with the quenched GT strengths is reported in  Table \ref{tab:GT}.

In the next section the theoretical results corresponding to $^{60}$Ge($0^+$) $\rightarrow$ $^{60}$Ga($1_f^+$)  
and $^{62}$Ge($0^+$) $\rightarrow$ $^{62}$Ga($1_f^+$) transitions are compared with the experimental data as reported in Ref. \cite{Orrigo}.

 \begin{figure}[h]
\begin{center}
	\includegraphics[width=6.8cm,height=4.5cm]{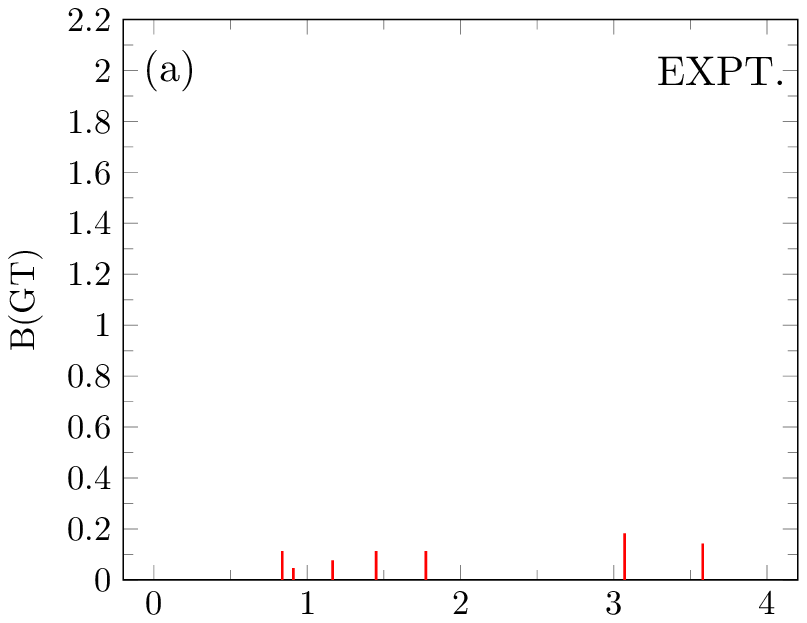}
	\includegraphics[width=6.8cm,height=4.5cm]{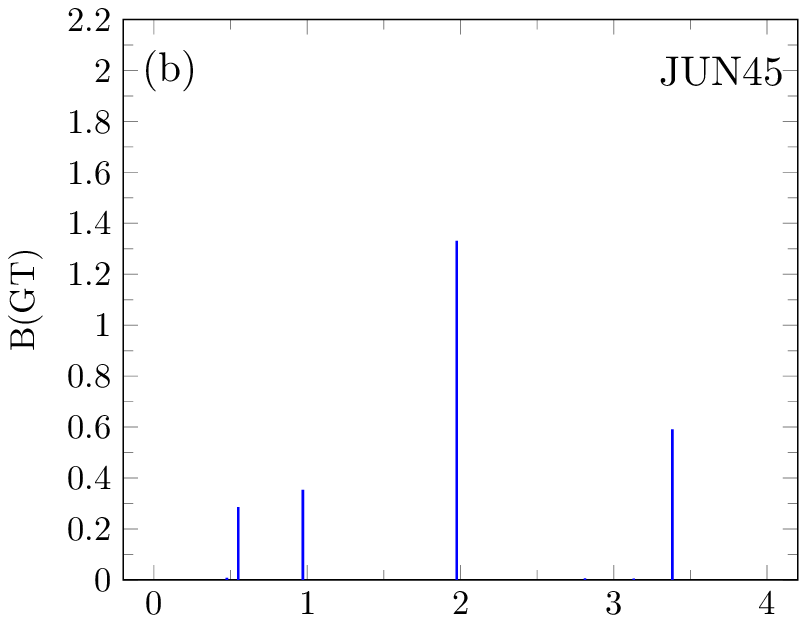}
	\includegraphics[width=6.8cm,height=4.5cm]{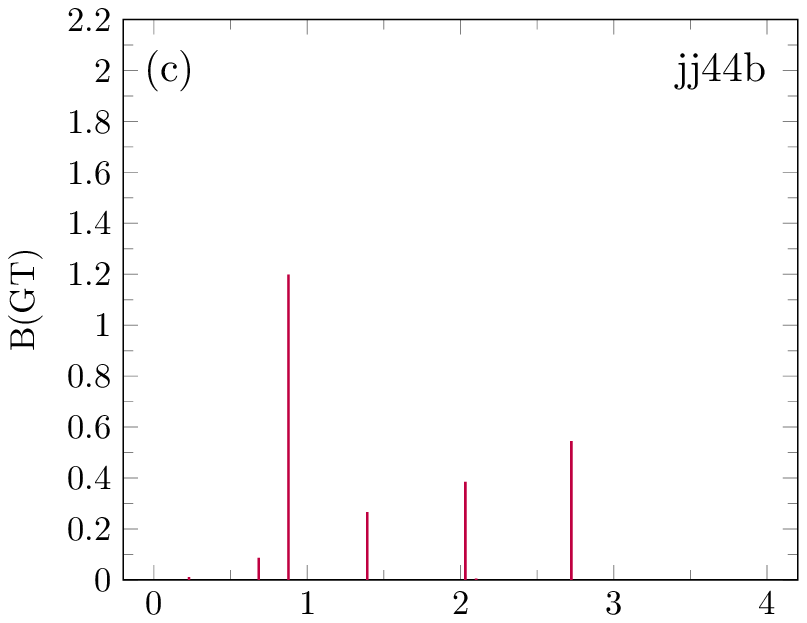}
	\includegraphics[width=6.8cm,height=4.5cm]{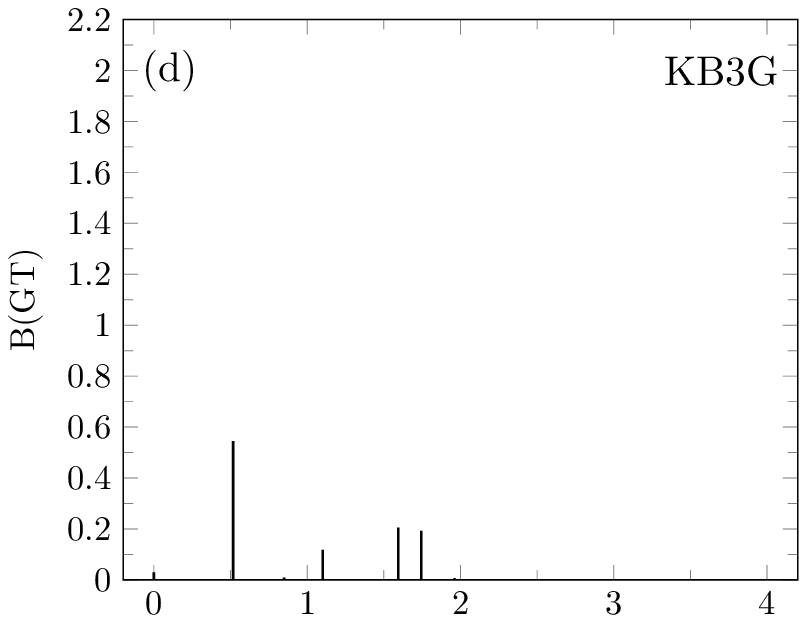}
	\includegraphics[width=6.8cm,height=4.5cm]{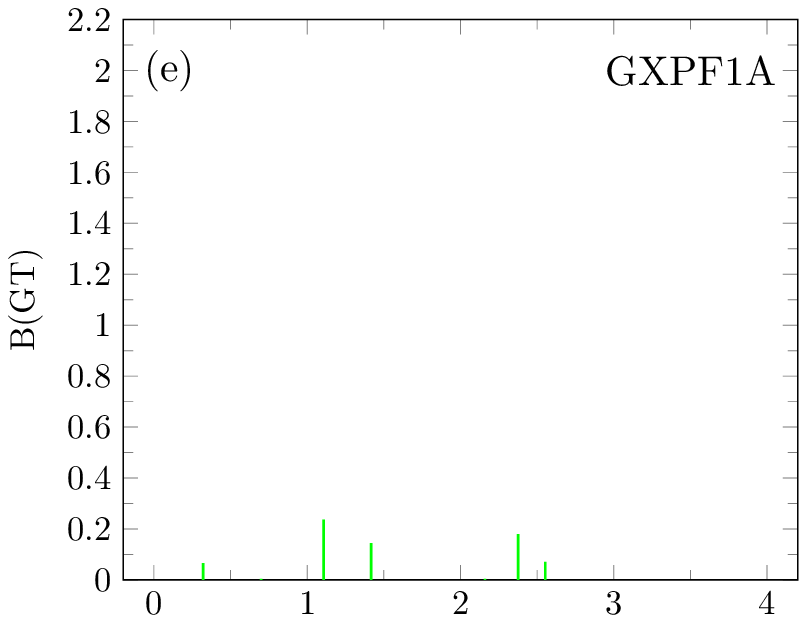}
	\includegraphics[width=6.8cm,height=4.7cm]{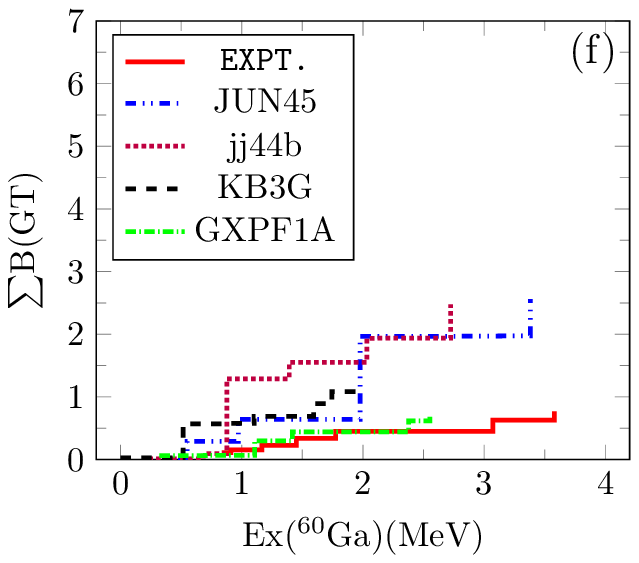}
	\caption{Comparison of GT-strengths between theory and experiment for $^{60}$Ge ($0^+$) $\rightarrow$ $^{60}$Ga($1^+$) transitions.}
\label{60GT}
\end{center}
\end{figure}

\begin{figure}[h]
\begin{center}
\includegraphics[width=6.8cm,height=4.5cm]{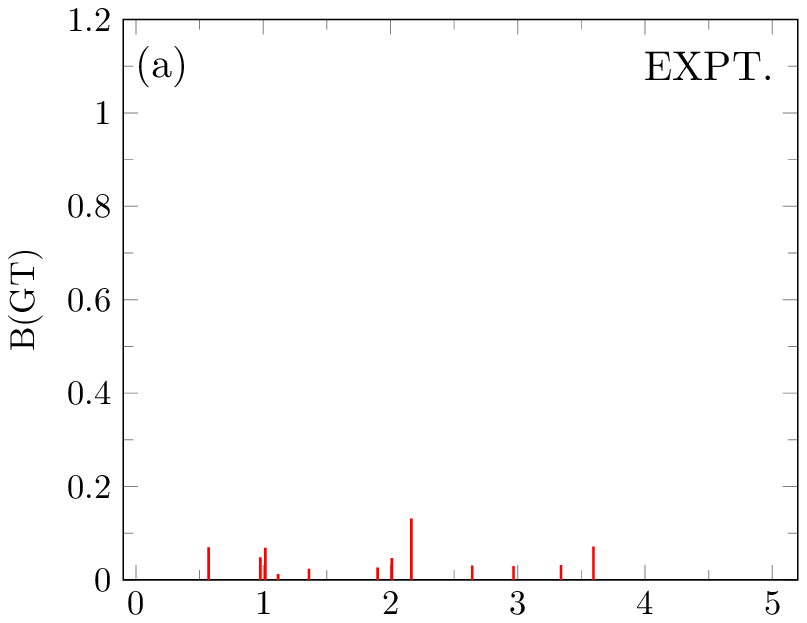}
\includegraphics[width=6.8cm,height=4.5cm]{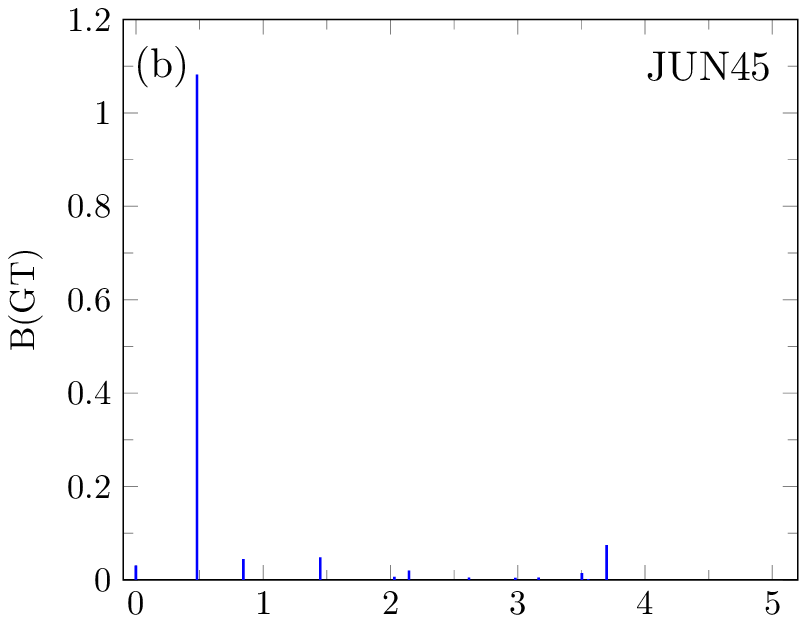}
\includegraphics[width=6.8cm,height=4.5cm]{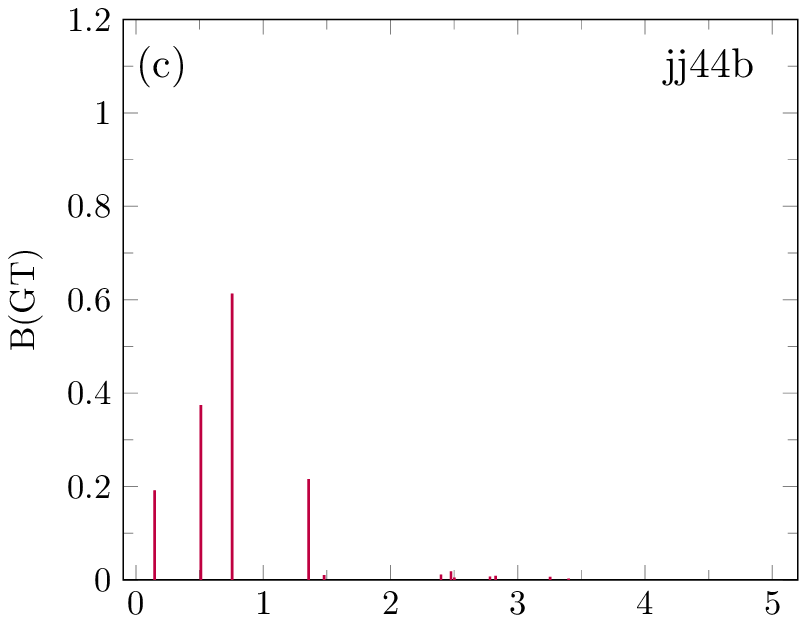}
\includegraphics[width=6.8cm,height=4.5cm]{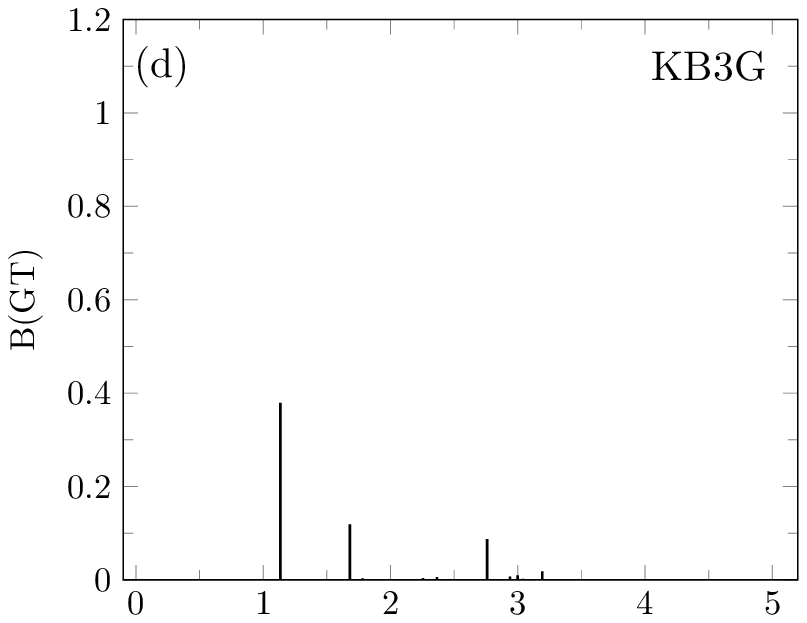}
\includegraphics[width=6.8cm,height=4.5cm]{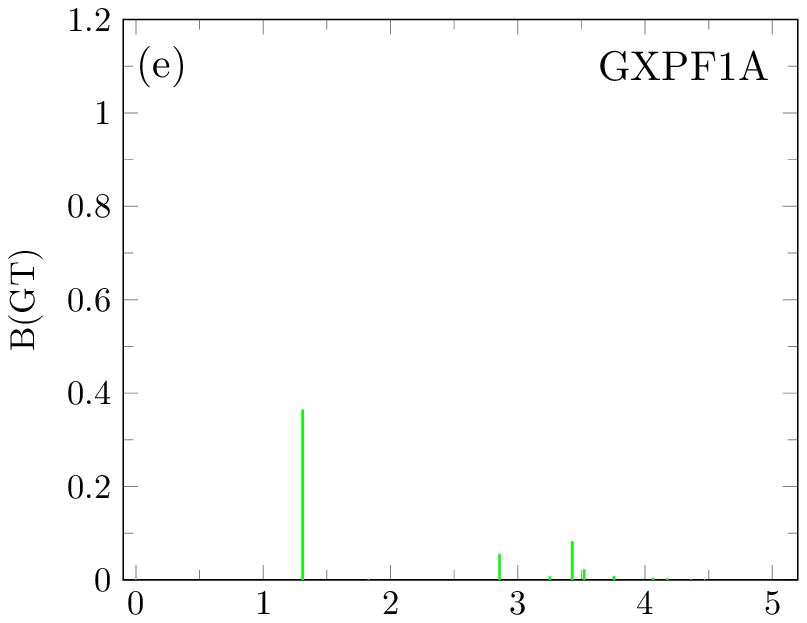}
\includegraphics[width=6.8cm,height=4.5cm]{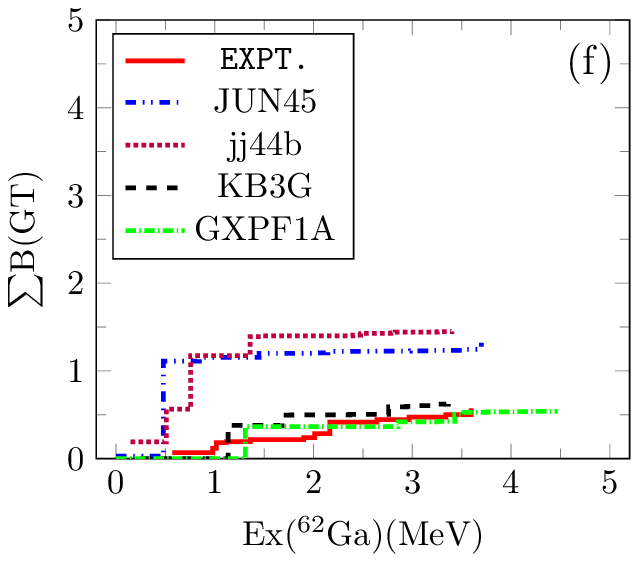}
\caption{Comparison of GT-strengths between theory and experiment for $^{62}$Ge($0^+$)  $\rightarrow$ $^{62}$Ga ($1^+$) transitions.}
\label{62GT}
\end{center}
\end{figure}

\subsection{{\bf GT-strengths corresponding to $^{60}$Ge($0^+$) $\rightarrow$ $^{60}$Ga($1_f^+$) transitions}}

Fig. \ref{60GT} displays a comparison between the shell-model calculations and the experimental GT strength
 distribution for the transition $^{60}$Ge $\rightarrow$ $^{60}$Ga. Fig. ~\ref{60GT}(a) presents the experimental data observed through the
$\beta$-decay $^{60}$Ge$\rightarrow ^{60}$Ga up to the excitation energy $E_x(^{60}$Ga) = 3.580 MeV \cite{Orrigo}.
Fig. ~\ref{60GT}(b) depicts the shell-model calculation using the JUN45 interaction, Fig. ~\ref{60GT}(c), the shell-model calculation using the jj44b interaction,
Fig. ~\ref{60GT}(d), the shell-model calculation using the KB3G interaction, Fig. ~\ref{60GT}(e), the shell-model calculation using the GXPF1A interaction, Fig. ~\ref{60GT}(f), the running sums of B(GT) as a function of the excitation energy. 

The seven experimental GT  transitions are observed corresponding to $^{60}$Ge$(0^+)$ $\rightarrow$ $^{60}$Ga(${1_f}^+$) transitions at 0.837, 0.910, 1.166, 1.450, 1.774, 3.071, and 3.580 MeV. The calculated GT strengths using JUN45 and jj44b interactions in $f_{5/2}pg_{9/2}$ space are comparatively larger than $fp$ model space as well as experimental ones. 
It is noticed that the effective interactions  JUN45 and jj44b generated excitation energy closer to the experimental one than the energy obtained employing the KB3G and GXPF1A interactions, while the opposite is true for the GT strength. 
The KB3G effective interaction predicts ${1^+}$ as a ground state and ${2^+}$ as a first excited state  which is 204 keV higher than $1^+$, while the order of the energy levels using GXPF1A interaction is exactly matching with the experiment. The calculated energy levels in the shell-model using KB3G and GXPF1A interactions are compressed as compared to the experimental ones this is because of the truncation imposed while filling the protons/neutrons in the model space, the truncation was necessary due to the computational limitations.
      Both the calculated GT strengths using KB3G and GXPF1A interactions in $fp$ space are  closer to experiments than the calculated results using JUN45 and jj44b interactions in $f_{5/2}pg_{9/2}$ space. This indicates that the $f_{7/2}$ orbital plays an important role in the calculations of B(GT) strengths for the  $^{60}$Ge $\rightarrow$ $^{60}$Ga transition. The close similarity in the B(GT) strength predicted using the GXPF1A interaction  is visible in the summed strength plot.


\subsection{{\bf GT-strengths corresponding to $^{62}$Ge($0^+$) $\rightarrow$ $^{62}$Ga($1_f^+$) transitions}}
  The comparison between the shell-model calculations and the experimental GT strength
 distribution for the  $^{62}$Ge $\rightarrow$ $^{62}$Ga transition are displayed in Fig. ~\ref{62GT}. Fig. ~\ref{62GT}(a) presents the experimental data observed through the
$\beta$-decay of $^{62}$Ge$\rightarrow ^{62}$Ga up to the excitation energy $E_x(^{62}$Ga) = 3.595 MeV \cite{Orrigo}.
Fig. ~\ref{62GT}(b) depicts the shell-model calculation using the JUN45 interaction, Fig. ~\ref{62GT}(c), the shell-model calculation using the jj44b interaction,
Fig. ~\ref{62GT}(d), the shell-model calculation using the KB3G interaction, Fig. ~\ref{62GT}(e), the shell-model calculation using the GXPF1A interaction, Fig. ~\ref{62GT}(f), the running sums of B(GT) as a function of the excitation energy. 

  The twelve GT strengths are observed corresponding to $^{62}$Ge $\rightarrow$ $^{62}$Ga transitions between 0.571 - 3.595 MeV. At lower excitation energies, the JUN45 and jj44b effective interactions reported larger B(GT) values than the experimental data, while beyond $\sim$ 1.4 MeV the calculated results are closer to the experimental data.
 Similar to the $^{60}$Ge $\rightarrow$ $^{60}$Ga transition, the KB3G and GXPF1A interactions generated GT strengths closer to the experimental one than the GT strengths obtained employing the JUN45 and jj44b interactions, while the opposite is true for the excitation energy.
 The JUN45 and GXPF1A interactions are predicting $1^+$ as a ground state and $0^+$ as a first excited state. The difference between these two states is 5 and 352 keV in JUN45 and GXPF1A interactions, respectively. The truncated energy levels using KB3G interaction are better than GXPF1A interaction, while the overall GT strengths and summed B(GT) using GXPF1A interaction are  better than KB3G interaction.
The GT strengths from shell-model in $fp$ model space are closer to  the experimental data which reflects the importance of $f_{7/2}$ orbital.
The sum of the B(GT) strength predicted by KB3G and GXPF1A interactions are similar to the experimental data as shown in  Fig. ~\ref{62GT}(f). 
The predicted half-lives using GXPF1A for $^{60}$Ge and $^{62}$Ge are very large.

\begin{sidewaystable}
\hspace{-2cm}
	\begin{center}
		\caption{\label{tab:FBGT1} Results of the $pf$-shell SM calculation using the GXPF1A interaction.  The matrix elements M(GT) of GT transitions corresponding to $^{60}$Ge ($0^+$) $\rightarrow$ $^{60}$Ga($1_f^+$) transitions are shown for each configuration. The results are shown for all excited GT states predicted in the region up to 2.554 MeV.
		The notation f7 $\rightarrow$ f7 stands for the transition with the $\nu$$f_{7/2}$$\rightarrow$$\pi$$f_{7/2}$ type. The summed value of the matrix elements is denoted by $\Sigma$ M(GT) and its squared value is the B(GT). In the last column, we have shown results corresponding to the  quenched value.}
		\scalebox{0.95}{%
\begin{tabular}{| p {1.8 cm} | p {1.3 cm} | p {1.3 cm} | p {1.3 cm} | p {1.3 cm} | p {1.3 cm} | p {1.3 cm} | p {1.3 cm} | p {1.3 cm} | p {1.3 cm} | p {1.3 cm} | p {1.3 cm} | p {1.3 cm} | p {1.12 cm} |p {1.9 cm} |p {1.9cm} |}
			\hline
			\hline
			\multicolumn{1} {|c|} {$^{60}Ge$ ($1^+$)}  &   \multicolumn{10} {c|} {Configurations}   & \multicolumn{3} {|c|} {} \\
			\cline{1-2}
			\cline{3-7}
			\cline{8-14}
			$E_x$ (MeV) &   f7$\rightarrow$f7& f7$\rightarrow$f5 &  p3$\rightarrow$p3 & p3$\rightarrow$f5 & p3$\rightarrow$p1 &f5$\rightarrow$f7 & f5$\rightarrow$p3 & f5$\rightarrow$f5&  p1$\rightarrow$p3  &  p1$\rightarrow$p1 & {\small{$\Sigma$ M(GT)}} &  B(GT) & $q^2$ x B(GT) \\
			\hline  
        0.321   &   0.0067 & -0.1736 & 0.6391 & 0.0000 & -0.0154 & -0.3208 & 0.0000 & -0.0754 & 0.3621 & -0.0407   & 0.3820  & 0.1459 &  0.0636 \\
		0.700   &   0.0104 & 0.0860 & 0.2193 & 0.0000 & -0.6783 & -0.0272 & 0.0000 & 0.1103 & 0.1883 & 0.0403     & 0.0500  & 0.0025 & 0.0010\\
		1.107	&   0.0549 & -0.2544 & 1.1126 & 0.0000 & 0.0456 & -0.1373 & 0.0000 & -0.0871 & -0.0073 & 0.0058   & 0.7328  & 0.5371 & 0.2339 \\
		1.417	&   0.0314 & -0.3385 & -0.1860 & 0.0000 & 0.8166 & -0.2786 & 0.0000 & 0.1355 & 0.4024 & -0.0132    & 0.5695  & 0.3244 & 0.1413 \\
		2.160	&  -0.0005 & 0.0909 & -0.0481 & 0.0000 & 0.0971 & 0.0449 & 0.0000 & -0.4118 & 0.1932 & -0.0209     & 0.05477  & 0.0030 & 0.0013\\
		2.376	&   0.0190 &-0.4268 & 0.3875 & 0.0000 & 0.3629 & -0.1324 & 0.0000 & 0.3358 & 0.0995 & -0.0085    & 0.6371  & 0.4059 &  0.1767 \\
		2.554	&  -0.0561& 0.1439 & -0.2656 & 0.0000 & 0.1807 & 0.0662 & 0.0000 & -0.2830 & -0.0378 & -0.1446   &0.3963   &0.1571 & 0.0685 \\
\hline			
			
		\end{tabular}}
	\end{center}
\hspace{-1cm}
	\begin{center}
		\caption{\label{tab:FBGT2} 
		Same as Table \ref{tab:FBGT1}, but for KB3G effective interaction in the case of $^{60}$Ge ($0^+$) $\rightarrow$ $^{60}$Ga ($1_f^+$) transitions.
			}
		\scalebox{0.95}{%
		\begin{tabular}{| p {1.8 cm} | p {1.3 cm} | p {1.3 cm} | p {1.3 cm} | p {1.3 cm} | p {1.3 cm} | p {1.3 cm} | p {1.3 cm} | p {1.3 cm} | p {1.3 cm} | p {1.3 cm} | p {1.3 cm} | p {1.3 cm} | p {1.12 cm} |p {1.9 cm} |p {1.9cm} |}
			\hline
			\hline
			\multicolumn{1} {|c|} {$^{60}Ge$ ($1^+$)}  &   \multicolumn{10} {c|} {Configurations} & \multicolumn{3} {|c|} {}    \\
			\cline{1-2}
			\cline{3-7}
			\cline{8-14}
			$E_x$ (MeV) &   f7$\rightarrow$f7& f7$\rightarrow$f5 &  p3$\rightarrow$p3 & p3$\rightarrow$f5 & p3$\rightarrow$p1 &f5$\rightarrow$f7 & f5$\rightarrow$p3 & f5$\rightarrow$f5&  p1$\rightarrow$p3  &  p1$\rightarrow$p1 & $\Sigma$ M(GT) &  B(GT) & $q^2$ x B(GT) \\
			\hline  
           0.000 &   0.0097& 0.0661& -0.2823 & 0.0000 & -0.2189 & 0.1222 & 0.0000 & 0.1719 & -0.1352 & 0.0199   & 0.2466  & 0.0608 & 0.0265\\
		   0.517 &  0.0177 & -0.2810 & 0.7486 & 0.0000 & 0.6124 & -0.1821 & 0.0000 & -0.0334 & 0.2453 & -0.0124  & 1.1152  & 1.2436 & 0.5423\\
			0.850 & 0.0012 & -0.0149 & 0.5104 & 0.0000 & -0.5272 & -0.1518 & 0.0000 & 0.0673 & 0.2405 & -0.0024   & 0.1232  &  0.0152& 0.0066\\
			1.102 &  -0.0300 & 0.0590 & -0.8177 & 0.0000 & 0.1395 & -0.0502 & 0.0000 & 0.1576 & 0.0516 & -0.0240   & 0.5144  & 0.2646 & 0.1153\\
			1.594 & -0.0146 &0.1742 & -0.0122 & 0.0000 & -0.5397 & 0.1552 & 0.0000 & -0.0094 & -0.4379 & 0.0024   & 0.6819  & 0.4650 & 0.2024\\
			1.744 &  -0.0249 & 0.2527 & -0.2277 & 0.0000 & -0.0169 & 0.1304 & 0.0000 & -0.7449 & -0.0137 & -0.0145   &  0.6596 & 0.4351 & 0.1894 \\
			1.961 &  0.0008 & 0.0626 & 0.0119 & 0.0000 & -0.2311 & -0.0316 & 0.0000 & 0.0673 & 0.1987 & 0.0262   &  0.10148 & 0.0110 & 0.0048\\
\hline			
			
		\end{tabular}}
	\end{center}
\end{sidewaystable}

   \begin{sidewaysfigure}
\begin{center}
	\includegraphics[width=20cm,height=12.5cm]{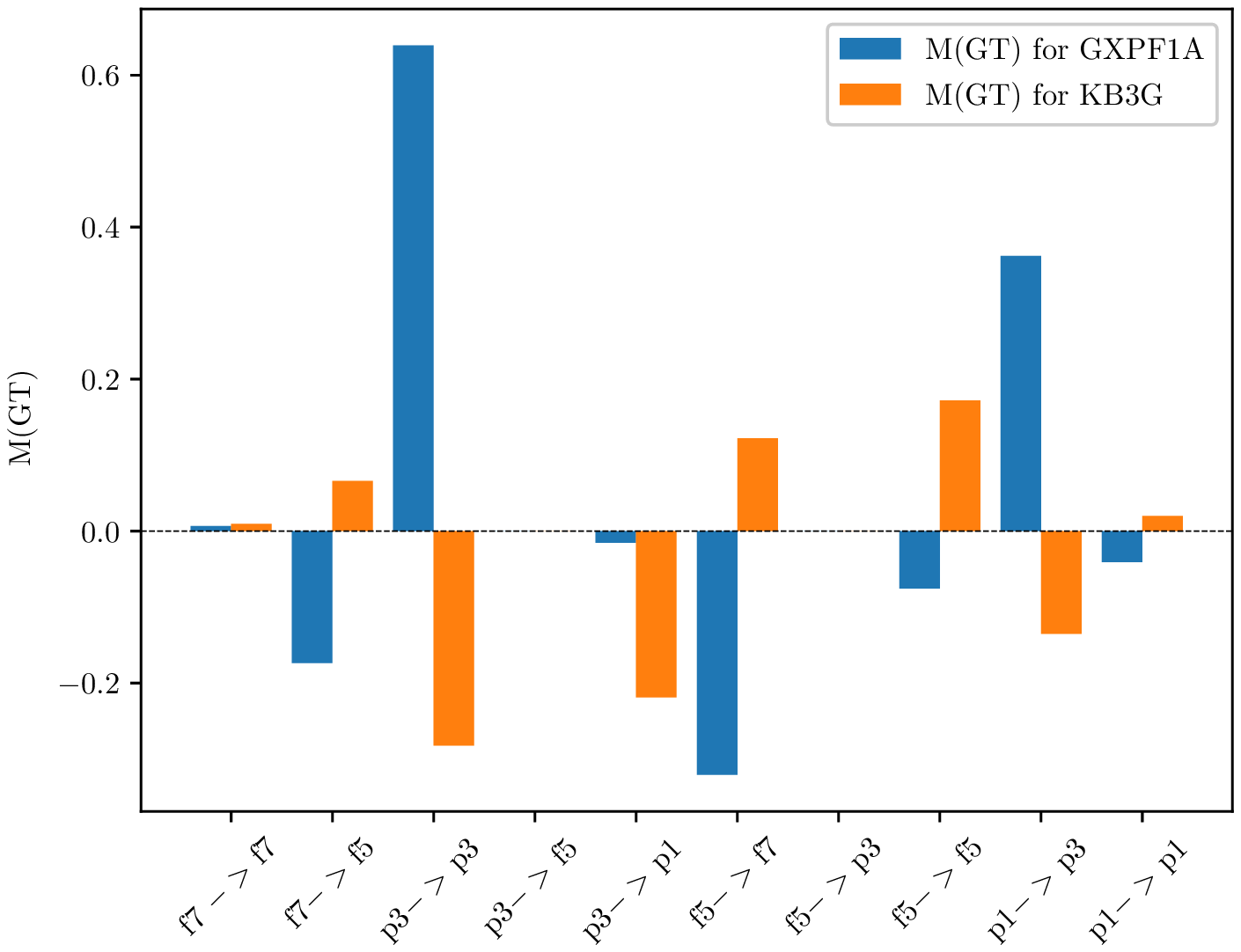}
	\caption{Comparison of GT-strength contribution from different orbitals between GXPF1A and KB3G for $^{60}$Ge ($0^+$) $\rightarrow$ $^{60}$Ga ($1_1^+$) transition.}
\label{distribution}
\end{center}
\end{sidewaysfigure}


\subsection{{\bf Different orbitals contribution to GT strengths}}

In the Tables \ref{tab:FBGT1}-\ref{tab:FBGT2}, the contributions from the different orbitals in the total M(GT)-strengths corresponding to $^{60}$Ge for KB3G and GXPF1A are shown. For GXPF1A, the lowest $1^+$ state predicted at 0.321 MeV, we see that the matrix elements of the configurations $\nu f_{7/2}$$\rightarrow$$\pi f_{7/2}$,
$\nu p_{3/2}$$\rightarrow$$\pi p_{3/2}$ and $\nu p_{1/2}$$\rightarrow$$\pi p_{3/2}$ are in the phase. In the case of KB3G interaction for the lowest $1^+$ state (see table \ref{tab:FBGT2}),  the contribution from $\nu f_{7/2}$$\rightarrow$$\pi f_{7/2}$,
$\nu f_{7/2}$$\rightarrow$$\pi f_{5/2}$, $\nu f_{5/2}$$\rightarrow$$\pi f_{7/2}$, $\nu f_{5/2}$$\rightarrow$$\pi f_{5/2}$  and  $\nu p_{1/2}$$\rightarrow$$\pi p_{1/2}$ are in the phase. It is clear that the role of $f_{7/2}$ orbital is important, it is contributing significantly to the total M(GT)-strengths.  In  Fig. \ref{distribution}, we have shown a comparison of GT-strength contribution from different orbitals between GXPF1A and KB3G for $^{60}$Ge ($0^+$) $\rightarrow$ $^{60}$Ga($1_1^+$) transition.

\section{Summary and Conclusions}\label{conclusions}

In the present work, we have performed shell-model calculations for GT-strengths
corresponding to recently available experimental data for $^{60}$Ge
$\rightarrow $ $^{60}$Ga and $^{62}$Ge $\rightarrow $ $^{62}$Ga transitions.
To know the importance of $f_{7/2}$ and $g_{9/2}$ orbitals we have performed
calculations in two different model spaces. For $fp$ model space we have
used KB3G and GXPF1A interactions, while for $f_{5/2}pg_{9/2}$ model space
we have used JUN45 and jj44b effective interactions. The calculated results
are in a reasonable agreement with the $fp$ model space. To show the importance
of different orbitals, we have also shown a contribution of different orbitals
in the calculated M(GT) values. Present theoretical work will add more
information to Ref.~\cite{Orrigo} where experimental data are reported.

\section*{Acknowledgments}
V.K. acknowledges financial support from SERB Project (EEQ/2019/000084), Govt. of India. 
The support and the resources provided by PARAM Shivay Facility under the National Supercomputing Mission, Government of India at the Indian Institute of Technology, Varanasi are gratefully acknowledged. P.C.S. acknowledges a research grant from SERB (India), CRG/2019/000556.
We would like to thank  Prof. Y. Fujita and  Dr. S. E. A. Orrigo for useful discussions during this work.



\begin{thebibliography}{44}
\expandafter\ifx\csname natexlab\endcsname\relax\def\natexlab#1{#1}\fi
\expandafter\ifx\csname bibnamefont\endcsname\relax
  \def\bibnamefont#1{#1}\fi
\expandafter\ifx\csname bibfnamefont\endcsname\relax
  \def\bibfnamefont#1{#1}\fi
\expandafter\ifx\csname citenamefont\endcsname\relax
  \def\citenamefont#1{#1}\fi
\expandafter\ifx\csname url\endcsname\relax
  \def\url#1{\texttt{#1}}\fi
\expandafter\ifx\csname urlprefix\endcsname\relax\def\urlprefix{URL }\fi
\providecommand{\bibinfo}[2]{#2}
\providecommand{\eprint}[2][]{\url{#2}}
	

	
\bibitem{stroberg} S. R. Stroberg, J. D. Holt, A. Schwenk, and J. Simonis,
{ \color{magenta} Phys. Rev. Lett. {\bf 126}, 022501 (2021).}
	
\bibitem{nunes} F. Nunes, 
{ \color{magenta} Physics Today {\bf 74}, 5, 34 (2021).}
	
\bibitem{otsuka} T. Otsuka, A. Gade, O. Sorlin, T. Suzuki, and Y. Utsuno, 
{\color{magenta} Rev. Mod. Phys. {\bf 92}, 015002 (2020).}
	
\bibitem{fujita} Y. Fujita, B. Rubio and W. Gelletly,
{\color{magenta} Prog. Part. Nucl. Phys. {\bf 66}, 549 (2011).}

\bibitem{jouni_review} H. Ejiri,  J. Suhonen and K. Zuber, 
{\color{magenta} Phys. Rep. {\bf 797}, 1 (2019).}

\bibitem{Orrigo1} S. E. A. Orrigo et al., 
{\color{magenta} Phys. Rev. Lett. {\bf 112}, 222501 (2014).}

\bibitem{Orrigo2} S. E. A. Orrigo et al.,  
{\color{magenta} Phys. Rev. C {\bf 93}, 044336 (2016).}

\bibitem{Molina} F. Molina et al.,
{\color{magenta} Phys. Rev. C {\bf 91}, 014301 (2015).}

\bibitem{Adachi1} T. Adachi et al.,
{\color{magenta} Phys. Rev. C {\bf 85}, 024308 (2012).}

\bibitem{Adachi2} T. Adachi et al.,
{\color{magenta} Phys. Rev. C {\bf 73}, 024311 (2006).}


\bibitem{Ciemny2016}
A. A. Ciemny et al., 
{\color{magenta} Eur. Phys. J. A {\bf 52}, 89 (2016).}

\bibitem{Lopez2001}
M. J. L$\acute{\rm o}$pez Jim$\acute{\rm e}$nez et al.
{\color{magenta} Phys. Rev. C {\bf 66}, 025803 (2002).}

\bibitem{Mazzocchi2001}
C .Mazzocchi et al. 
{\color{magenta} Eur. Phys. J. A {\bf 12}, 269 (2001).}

\bibitem{Grodner} E. Grodner et al., 
{\color{magenta} Phys. Rev. Lett. {\bf 113}, 092501 (2014).}

\bibitem{Kucuk2017}
L. Kucuk et al.,
{\color{magenta}Eur. Phys. J. A {\bf 53}, 134 (2017). }

%
%

\bibitem{Aberg} A. Juodagalvis and S. Aberg, 
{\color{magenta} Nucl. Phys. A {\bf 683}, 207  (2001).}

\bibitem{PCS} P.C. Srivastava, R. Sahu and V.K.B. Kota, 
{\color{magenta} Eur. Phys. J. A {\bf 51}, 3  (2015).}

\bibitem{Pittel} S. Pittel, A. Carranza M., and J. G. Hirsch, 
{\color{magenta} Journal of Physics: Conference Series {\bf 1610}, 012012 (2020).}

\bibitem{Orrigo} S.E.A. Orrigo et al., 
{\color{magenta} Phys. Rev. C {\bf 103}, 014324 (2021).}

\bibitem{wilkinson} D. H. Wilkinson and B. E. F. Macefield, {\color{magenta} Nucl. Phys. A {\bf 232}, 58 (1974).}

\bibitem{Sirlin} A. Sirlin and R. Zucchini, { \color{magenta} Phys. Rev. Lett. {\bf 57}, 1994 (1986).}

\bibitem{wga} D. H. Wilkinson, A. Gallmann, and D. E. Alburger, {\color{magenta} Phys. Rev. C {\bf 18}, 401 (1978).}

	
\bibitem{Caurier59}	
	E. Caurier, G. Mart\'{\i}nez-Pinedo, F. Nowacki, A. Poves, J. Retamosa, and A. P. Zuker,
	{\color{magenta} Phys. Rev. C {\bf 59}, 2033 (1999)}.
	

\bibitem{brownwild} B. A.  Brown and  B. H. Wildenthal,
{\color{magenta} At. Data Nucl. Data Tables \textbf{33}, 347 (1985).}

\bibitem{Mart} G. Martinez-Pinedo and A. P. Zuker,
{\color{magenta} Phys. Rev. C {\bf 53}, R2602(R) (1996).}



\bibitem{Archana2018} A. Saxena, P.C. Srivastava and T. Suzuki,
{\color{magenta} Phys. Rev. C {\bf {97}}, 024310 (2018)}.

\bibitem{Anil2020} A. Kumar, P.C. Srivastava and T. Suzuki, 
{\color{magenta} Prog. Theo. Expt. Phys. {\bf {2020}},  033D01 (2020)}.

\bibitem{Vikas2016} V. Kumar, P.C. Srivastava, and H. Li,
{\color{magenta} Jour. Phys. G: Nucl. and Part. Phys. {\bf 43}, 105104 (2016).}

\bibitem{Vikas22016} V. Kumar and P.C. Srivastava, 
{\color{magenta} Eur. Phys. J. A {\bf {52}}, 181 (2016).}

\bibitem{Anil2020PRC} A. Kumar, P. C. Srivastava, J. Kostensalo and J. Suhonen,
	{\color{magenta} Phys. Rev. C. {\bf 101}, 064304 (2020)}.
	
\bibitem{Anil2021EPJA} A. Kumar, P. C. Srivastava  and J. Suhonen,
	{\color{magenta} Eur. Phys. J. A,  {\bf 57}, 225 (2021)}.
	
\bibitem{PriyankaPRC} P. Choudhary, A. Kumar,  P.C. Srivastava, T. Suzuki,
	{\color{magenta} Phys. Rev. C {\bf 103}, 064325 (2021)  }.
	

	
\bibitem{vikasnpa} V. Kumar and P.C. Srivastava, 
	{\color{magenta} Nucl. Phys. A  {\bf 1002}, 121989 (2020)}.

\bibitem{jun45} M. Honma, T. Otsuka, T. Mizusaki and M. Hjorth-Jensen,
 {\color{magenta} Phys. Rev. C {\bf 80}, 064323 (2009).}

\bibitem{jj44b} B.A. Brown and A.F. Lisetskiy (unpublished); see also endnote
(28) in B. Cheal {\it et al.}, 
{\color{magenta} Phys. Rev. Lett. {\bf 104}, 252502 (2010).}

\bibitem{KB3G} E. Caurier, K. Langanke, G. Martínez-Pinedo, F. Nowacki,
{ \color{magenta}Nucl. Phys. A {\bf 653}, 439 (1999).}

\bibitem{GXPF1A} M. Honma $\it et ~ al.$, 
{\color{magenta} Eur. Phys. J. A {\bf 25}, 499 (2005).}

\bibitem{KB3} A. Poves and A. Zuker,
{ \color{magenta} Phys. Rep. {\bf 70}, 235 (1981).}

\bibitem{GXPF1} M. Honma $\it et ~ al.$, 
{ \color{magenta} Phys. Rev. C {\bf 69}, 034335 (2004).}

\bibitem{NuShellX@MSU} B. A. Brown  $\it et ~ al.$, 
{\color{magenta} Nucl. Data Sheets {\bf 120}, 115 (2014).}

\bibitem{kshell} N. Shimizu $\it et ~ al.$, 
{\color{magenta} Comp. Phys. Com. {\bf 244}, 372 (2019).}
\end{thebibliography}
\end{document}